\documentclass[prb,preprintnumbers,superscriptaddress,shownopacs,twocolumn,balancelastpage,floatfix]{revtex4}
\usepackage{amsmath,amssymb,amsfonts}
\usepackage{bm,graphicx,color}

\newcommand{\TR}{\text{Tr}}

\newcommand{\tb}{{\bar t}}

\newcommand{\mydagger}{{\dagger}}
\newcommand{\phdagger}{{\phantom{\dagger}\!}}
\newcommand{\bra}[1]{\langle{#1}|}
\newcommand{\ket}[1]{|{#1}\rangle}

\newcommand{\expval}[1]{\langle{#1}\rangle}
\newcommand{\CS}{\mathcal{S}}

\newcommand{\delC}{\delta_\CC}
\newcommand{\intC}{\int_\CC}
\newcommand{\tmin}{t_{\text{min}}}
\newcommand{\tmax}{t_{\text{max}}}
\newcommand{\CC}{\mathcal{C}}
\newcommand{\TC}{\text{T}_{\CC}}

\newcommand{\convz}{\ast}

\newcommand{\mat}{{\text{\tiny M}}}

\newcommand{\YY}{Y}
\newcommand{\KK}{K}
\newcommand{\pseudo}{\widetilde}
\newcommand{\cyles}{\prec}
\newcommand{\scell}{\text{scel}}
\newcommand{\pG}{\mathcal{G}}

\begin{document}

  \title{Nonequilibrium dynamical mean-field calculations based on the non-crossing 
  approximation and its generalizations}

  \author{Martin Eckstein}
  %\affiliation{Theoretical Physics, ETH Zurich, 8093 Zurich, Switzerland}

  \author{Philipp Werner}
  \affiliation{Theoretical Physics, ETH Zurich, 8093 Zurich, Switzerland}  
  
  \date{\today}

\begin{abstract}
We solve the impurity problem which arises within nonequilibrium dynamical mean-field theory 
for the Hubbard model by means of a self-consistent perturbation expansion around the atomic 
limit. While the lowest order, known as the non-crossing approximation (NCA), is reliable only 
when the interaction $U$ is much larger than the bandwidth, low-order corrections to the NCA 
turn out to be sufficient to reproduce numerically exact Monte Carlo results in a wide parameter 
range that covers the insulating phase and the metal-insulator crossover regime at not too 
low temperatures. As an application of the perturbative strong-coupling impurity solver we 
investigate the response of the double occupancy in the Mott insulating phase of the Hubbard 
model to a dynamical change of the interaction or the hopping, a technique which has been 
used as a probe of the Mott insulating state in ultracold fermionic gases.
\end{abstract}

\maketitle

\section{Introduction}

Experiments with ultracold atoms in optical lattices,\cite{Bloch2008a} as well
as pump-probe spectroscopy with femtosecond time-resolution%
\cite{Iwai2003a,Perfetti2006a,Wall2009} and transport measurements on quantum dots%
\cite{Goldhaber1998} enable a systematic investigation of strongly interacting quantum
many-particle systems under nonequilibrium conditions. In a pump-probe experiment, a
material is excited with a strong laser pulse, and its subsequent evolution is probed
by a second pulse that reaches the sample at a controlled time-delay. The breakdown of the Mott insulating
phase within a few times of the inverse hopping has been observed in this way.%
\cite{Perfetti2006a,Wall2009} Ultracold gases in optical lattices, on the other hand, can
be prepared in an equilibrium state and suddenly quenched out of equilibrium by modifying 
a Hamiltonian parameter.\cite{Greiner2002b} These experiments can address fundamental 
questions such as the thermalization in isolated quantum systems.\cite{Kinoshita2006}

The ongoing experimental progress has stimulated intensive research on the theoretical side. 
While many theoretical approaches that are designed for the investigation of correlated 
systems in thermal equilibrium must be modified considerably before they can be 
used to compute the real-time evolution, dynamical mean-field theory (DMFT)%
\cite{Georges1996} is an approximate scheme which is per se applicable to both 
equilibrium and nonequilibrium situations.\cite{Schmidt2002} The method relies on a 
mapping of lattice models to a single impurity model, which is exact in the limit of 
infinite dimensions,\cite{Metzner1989a} and provides a good basis for the realistic 
simulation of many correlated materials.\cite{Kotliar2004} Nonequilibrium DMFT 
has so far been used, e.g., to study transport beyond linear response in the 
Falicov-Kimball model,\cite{Freericks2006a,Freericks2008b,Tsuji2008a,Tsuji2009a}
as well as interaction quenches and interaction ramps in the Falicov-Kimball model%
\cite{Eckstein2008a,Eckstein2009b} and in the Hubbard model.% 
\cite{Eckstein2009a,Eckstein2010a} In the last part of this paper we will use 
DMFT to study the response of the Mott insulating phase to a periodic modulation
of the hopping or the interaction,\cite{Kollath2006a} similar to what can now be done in 
experiments with cold atomic gases.\cite{Joerdens2008a,Strohmeier2010a} 

Currently, the biggest challenge within the context of nonequilibrium DMFT is the 
development of impurity solvers which allow to compute the long-time dynamics after a 
perturbation. An exact solution, via a closed set of equations of motion, is known 
only for the Falicov-Kimball model.\cite{Brandt1989,Freericks2006a} For the Hubbard 
model, continuous-time Quantum Monte Carlo (CTQMC) can in principle be used to obtain 
an unbiased solution.\cite{Eckstein2009a,Eckstein2010a} Both the weak-coupling expansion%
\cite{Werner2009a} and the strong-coupling expansion\cite{Schiro2009b} of the 
relevant Anderson impurity model have been translated from their respective 
imaginary-time variants (Refs.~\onlinecite{Rubtsov2005a} and \onlinecite{Gull2008a} 
for weak-coupling and Ref.~\onlinecite{Werner2006a} for strong-coupling) to 
the Keldysh formalism, in order to study the real-time evolution. However, in 
these real-time Monte Carlo calculations, the accessible times are limited 
by the notorious dynamical sign problem. A big advantage of the 
weak-coupling expansion over the strong-coupling expansion 
is that the diagrammatic series simplifies in the case of particle-hole symmetry.%
\cite{Werner2010a} On the other hand, the sign problem in a weak-coupling 
calculation increases with the interaction strength. It is thus essentially impossible 
to use CTQMC to study complex excitation processes within the Mott insulating phase, 
e.g., the excitation with a short laser pulse and the subsequent relaxation. 

In order to avoid the sign problem and access the regime of strong interactions and 
relatively long times, we explore the direct summation of the self-consistent 
diagrammatic hybridization expansion up to fixed order, as opposed to CTQMC, 
which is in essence a stochastic summation of the full (non self-consistent) 
series. This approach proves to be very accurate in a wide paremeter regime 
and suitable for the calculation of the real-time dynamics.

Systematic approximations for the expansion around the atomic limit of the 
Anderson impurity model have been used for a long 
time.\cite{Keiter1971a,Grewe1981a,Kuramoto1983a,Barnes1976a,Coleman1984a,
Bickers1987a,Bickers1987b} The simplest conserving approximation, which 
has been termed non-crossing approximation (NCA), %\cite{Kuramoto1983a}
can correctly recover the Kondo temperature $T_K$ when charge fluctuations 
are suppressed by the Coulomb interaction $U$, although the Fermi-liquid 
behavior for $T \ll T_K$ is not correctly reproduced.\cite{MuellerHartmann1984a,Bickers1987b} 
If $U$ is finite, however, the width of the Kondo resonance is severely 
underestimated, and various resummation schemes of the expansion have 
been devised to cure this problem.\cite{Pruschke1989a,Haule2001a} Already 
the simplest to NCA within these schemes, the so called one-crossing 
approximation (OCA), can cure the deficiencies of NCA to a large extent.
Motivated by the fact that NCA is already very good in the insulating parameter 
regime, Gull {\it et al.}\cite{Gull2010a} recently developed a bold-line
hybridization expansion, i.e., an approach which is based on a Monte Carlo 
sampling of the corrections beyond NCA.

Starting with the work of Pruschke, Cox, and Jarrell,\cite{Pruschke1993a}
both the NCA and the OCA have been used as an impurity solver for DMFT 
(for some recent references that involve the investigation of real 
materials, see Refs.~\onlinecite{Shim2007a,Shim2007b,Haule2009a}). 
Furthermore, NCA and its corrections can readily be translated into 
the Keldysh formalism to study nonequilibrium situations, although 
the evaluation of higher-order diagrams in real time involves quite 
some numerical effort. For example, the buildup of the Kondo resonance 
after a sudden shift of the impurity level in the Kondo model has been 
investigated with NCA.\cite{Nordlander1999a} The fairly accurate results 
in equilibrium calculations and the straightforward portability to the 
Keldysh contour make the self-consistent hybridization expansion an 
interesting candidate for the solution of the impurity problem 
within nonequilibrium DMFT.

The purpose of this paper is twofold. First, we give a detailed description 
of  the self-consistent expansion on the Keldysh contour (Sections.~\ref{sec-def}
and \ref{sec-nca}), and we benchmark the method by applying it to the interaction 
quench in the Hubbard model on the Bethe-lattice 
(Sec.~\ref{sec-qmccomparison}). We find similar trends for both equilibrium 
and nonequilibrium: While NCA is unreliable unless the interaction $U$ is much 
larger than the bandwidth, OCA provides an important correction, and the third 
order is in almost quantitative agreement with QMC results over a wide parameter 
range which includes the insulating phase and the crossover regime between metal 
and insulator.  As an application of the perturbative impurity solver 
we then study the excitation of a Mott insulator by a time-dependent modulation 
of the hopping or the interaction strength (Sec.~\ref{sec-spectroscopy}). 
Because the interaction is rather large in this problem, a solution using 
weak-coupling CTQMC is currently not feasible. 

\section{Definition of the impurity problem on the Keldysh contour}
\label{sec-def}

\begin{figure}
%   \centerline{\includegraphics[width=0.5\columnwidth]{Figures/figure01.eps}}
   \centerline{\includegraphics[width=0.5\columnwidth]{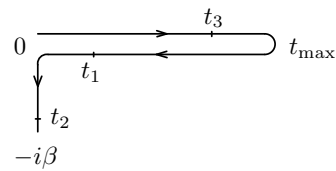}}
   \caption{
    The L-shaped contour $\CC$ for the description of transient 
    nonequilibrium states with initial state density matrix 
    $\propto \exp[-\beta H(0)]$. The indicated time arguments 
    are in cyclic order, $t_1 \cyles t_2 \cyles t_3$ (see text).}
    \label{fig-contour}
\end{figure}

To describe a nonequilibrium situation in which the system is prepared 
in a thermal equilibrium state at temperature $T=1/\beta$ for times 
$t<0$ and later acted on by some perturbation, we use the Keldysh 
formalism.\cite{Keldysh1964a,keldyshintro} The imaginary-time contour 
of the Matsubara Green's functions for finite-temperature equilibrium states 
is thereby extended to the L-shaped contour $\CC$ that runs from $0$ to time 
$t_\text{max}$ (i.e., the largest time of interest) along the real axis, back to 
$0$, and finally to $-i\beta$ along the imaginary time axis (Fig.~\ref{fig-contour}). 
The Keldysh formalism is based on the use of contour-ordered correlation 
functions $\expval{\TC A(t_1)B(t_2)}$, where $\TC$ exchanges the order of the 
two operators $A(t_1)$ and $B(t_2)$ in the 
product $A(t_1) B(t_2)$ if  $t_2$ appears later on the contour 
than $t_1$, according to the order which is indicated by the arrows in 
Fig.~\ref{fig-contour}. An additional minus sign appears if the exchange 
involves an odd number of Fermi operators. The use of contour-ordered Green's 
functions allows the application of Wick's theorem if the action 
is quadratic.\cite{Keldysh1964a} Depending on the choice of the time arguments, 
a contour-ordered correlation function describes either real-time correlations, 
or it recovers the imaginary-time ordered correlation function of the initial 
equilibrium state. 

In the following sections we consider an impurity model which is defined 
by the following action on the L-shaped contour $\CC$,
  \begin{subequations}
  \label{action}
  \begin{align}
  \CS_\text{imp} 
  &= 
  \CS_\text{loc} + \CS_\text{hyb},
  \\
  \label{sloc}
  \CS_\text{loc} 
  &= 
  -i\!
  \intC \!\!dt\, H_\text{loc}[d_{p}^\dagger(t),d_{p}(t),t],
  \\
  \label{shyb}
  \CS_\text{hyb}
  &=
  -i
  \!
  \intC \!\!dt_1 dt_2\,   
  \!\!
  \sum_{p_1,p_2}\!
  d_{p_1}^\dagger\!(t_1) \,\Lambda_{p_1,p_2}(t_1,t_2)\, d_{p_2}\!(t_2). 
  \end{align}
  \end{subequations}  
In this action, $d_{p}$ and $d_{p}^\dagger$ denote annihilation and creation operators for an electron 
in the impurity level $p$ ($p$ labels spin and orbital degrees of freedom), 
and $H_\text{loc}$ is the local Hamiltonian of the impurity site, which can be 
interacting and time-dependent in general. The hybridization function 
$\Lambda_{p_1,p_2}(t_1,t_2)$ gives the amplitude for the hopping of an 
electron from the $p_2$-orbital into the bath at time $t_2$, its 
propagation within the bath, and the hopping back into the impurity 
orbital  $p_1$ at time $t_1$. The action (\ref{action}) can be 
derived from an impurity Hamiltonian with time-dependent 
coupling between bath and impurity, 
\begin{equation}
H(t) 
=
H_\text{loc}(t) 
+ \sum_\nu \epsilon_\nu c_\nu^\dagger c_\nu 
+ \sum_{p,\nu} [\, V_{p,\nu}(t) \,d_p^\dagger c_\nu + h.c.\,] 
\end{equation}
by tracing out the bath degrees of freedom $c_\nu$. It also arises as the 
effective single-site problem in nonequilibrium DMFT,\cite{Freericks2006a} 
without direct reference to a given Hamiltonian formulation.

The single-particle Green's function of the impurity model (\ref{action}) is given 
by
\begin{equation}
  \label{impg}
    G_{p,p'}(t,t') = -i \expval{d_{p}(t)d_{p'}^\dagger (t')}_{\CS_\text{imp} },
\end{equation}
where the contour-ordered expectation value for the action $\CS$ is defined as
\begin{equation}
\label{cntr-expval}
  \expval{\cdots}_{\CS} = \frac{\TR [\TC \exp(\CS) \cdots]}
  {\TR [\TC \exp(\CS)]}.
\end{equation}
The nonequilibrium formalism presented below reduces to the Matsubara formalism 
for the initial equilibrium state when all calculations are restricted to the 
imaginary branch of the contour. For time-arguments $t=-i\tau$ and 
$t'=-i\tau'$ on the imaginary branch of $\CC$, the Green's function
$G(t,t')$ is directly related to the Matsubara Green's function $G^\mat(\tau)$ 
of the initial thermal equilibrium state,
\begin{equation}
\label{matsubara-def}
G(-i\tau,-i\tau') \equiv iG^\mat(\tau-\tau'),
\end{equation}
which is translationally invariant in time. An analogous equation is used 
to define the Matsubara component of all two-time contour-ordered correlation 
functions that are used in the following. The factor $i$ on the right-hand side 
of Eq.~(\ref{matsubara-def}) is needed to recover the conventional definition 
of the Matsubara functions.

\section{Self-consistent diagrammatic hybridization expansion 
of the Anderson Impurity model} 
\label{sec-nca}

  \subsection{Pseudoparticle representation}
    
  In this section we compute the Green's function (\ref{impg}) by expanding the 
  expectation value (\ref{cntr-expval}) in terms of the hybridization function 
  $\Lambda(t,t')$. The self-consistent hybridization expansion for the Anderson 
  model in thermal equilibrium has been described previously in many 
  places,\cite{Keiter1971a,Grewe1981a,Kuramoto1983a,Barnes1976a,Coleman1984a,%
  Bickers1987a,Bickers1987b,Pruschke1989a,Haule2001a,Haule2009a}
  %(for a recent reference, see, e.g., Ref.~\onlinecite{Haule2009a}),
  and the generalization from the imaginary-time contour to the Keldysh 
  contour is rather straightforward. Nevertheless we give a detailed 
  derivation below, in order to discuss some important technical differences 
  between the equilibrium and the nonequilibrium variants of the expansion.

  Because the local part (\ref{sloc}) of the action is generally not quadratic, 
  standard diagrammatic perturbation theory does not apply to the expansion 
  around the atomic limit ($\Lambda=0$). There exist several related strategies 
  to by-pass this difficulty. In the CTQMC variant of the hybridization expansion,%
  \cite{Werner2006a} high-order time-ordered correlation functions of the 
  impurity problem are explicitly evaluated in a suitable basis of the local 
  problem. We will follow a different approach, which is based on the introduction 
  of auxiliary particles.\cite{Barnes1976a,Coleman1984a} This allows to use 
  standard resummation tricks from diagrammatic perturbation theory, at the 
  expense of having to do a projection from the extended pseudoparticle 
  Hilbert space to the physical Hilbert space. 
  
  The local part $H_\text{loc}(t)$ of the impurity Hamiltonian is diagonalized 
  at each instant of time,
  \begin{equation}
  \label{hloc-diag}
  H_\text{loc}(t) = \sum_m \ket{m(t)} E_m(t) \bra{m(t)},
  \end{equation}
  and for each eigenstate $m$ one flavor of pseudoparticles, with 
  annihilation (creation) operator $a_m^{(\dagger)}$, is introduced.
  We assume that eigenstates $\ket{m(t)}$ are smooth functions of time. 
  Pseudoparticles are bosons if the state $\ket{m}$ corresponds to 
  an even number of particles on the impurity, and fermions otherwise. By 
  means of the isomorphy $\ket{m(t)} \leftrightarrow a_m^\dagger\ket{vac}$,
  the physical Hilbert space of the impurity can be identified with the 
  subspace of the pseudoparticle Fock space in which the total number 
  of pseudoparticles, 
  \begin{equation}
  \label{q}
  Q= \sum_m a_m^\dagger a_m,
  \end{equation}
  is exactly one. Hence the expectation value $\expval{A(t)}_{\CS_\text{imp}}$ of 
  any impurity observable $A$ can be computed in the pseudoparticle space as
  \begin{align}
  \label{trace1}
  \expval{A(t)}_{\CS_\text{imp}} &= \expval{\pseudo A(t)}_{Q=1}
  \equiv
  \frac{ \expval{\delta_{Q,1}\pseudo A(t)}_{\pseudo\CS_\text{imp}} }{\expval{\delta_{Q,1}}_{\pseudo\CS_\text{imp}}},
  \end{align}
  provided that the pseudoparticle action $\pseudo\CS_\text{imp}$ and the observable 
  $\pseudo A$ are constructed such that they coincide with  $\CS_\text{imp}$ and $A$
  in the $Q=1$ subspace ($\delta_{Q,1}$ is the projection onto $Q=1$). 
  The requirement is satisfied by choosing
  \begin{subequations}
  \label{pseudo-d}
  \begin{align}
  \pseudo d_{p}^\dagger(t)
  &= \sum_{m,n} F_{mn}^{\,p}\!(t)\, a_m^\dagger  a_n
  \\
  \pseudo d_{p}(t)
  &= \sum_{m,n} F_{nm}^{\,p}\!(t)^* \, a_m^\dagger  a_n
  \\
  \label{d-matrixelement}
  F_{mn}^{\,p}\!(t) 
  &=
  \bra{m(t)} d_{p}^\dagger \ket{n(t)}
  \end{align}
  \end{subequations}
  for the electron annihilation and creation operators, and  
  \begin{subequations}
  \label{pseudoaction}
  \begin{align}
  \pseudo \CS_\text{loc} 
  &= 
  -i \sum_m \intC \!dt\, 
  %[E_m(t)-\lambda_0] a_m^\dagger(t) a_m(t)
  E_m(t) \, a_m^\dagger(t) a_m(t)
  \\
  \pseudo \CS_\text{hyb}
  &=
  -i\sum_{m,n,m',n'} \sum_{p,p'} 
  \intC \!dt dt'\,   
  a_m^\dagger(t) a_{n}(t) \,\,\times
  \nonumber\\
  &\hspace*{0ex} \times\,\,
  F_{mn}^{\,p}\!(t)\, \Lambda_{p,p'}(t,t')\, {F_{n'm'}^{\,p'}(t)}^{\!*}
  a_{m'}^\dagger(t') a_{n'}(t')
  \end{align}
  \end{subequations}
  for the impurity action. The first line in Eq.~(\ref{pseudoaction}) follows 
  from Eqs.~(\ref{sloc}) and (\ref{hloc-diag}), and the second line results 
  from direct insertion of Eqs.~(\ref{pseudo-d}) into Eq.~(\ref{shyb}).
  
  Feynman diagrams for pseudoparticle propagators are most easily constructed 
  in the grand-canonical ensemble with respect to the total pseudoparticle 
  number $Q$. For this purpose it is convenient to switch into the interaction 
  representation with respect to a chemical potential term  
  $\lambda Q$. 
  Because $Q$ is a conserved quantity, the value of $\lambda$ for $t>0$ 
  has no influence on the expectation value of physical observables.
  We choose $\lambda$ to be present only at times $t<0$, (i.e., on the 
  imaginary part of the contour, where $t=-i\tau$), such that 
  \begin{subequations}
  \label{interactionrepresentation}
  \begin{align}
  a_m(t) &= 
  a_m\exp(\lambda\,\text{Im} t)
  \\
  a_m^{\dagger}(t) &= 
  a_m^{\dagger}\exp(- \lambda\,\text{Im} t),
  \end{align}
  \end{subequations}
  and the grand canonical average can be denoted as
  \begin{equation}
  \label{gge}
  \expval{\pseudo A(t)}_\lambda
  =
  \frac{ \expval{ e^{-\beta\lambda Q} \pseudo A(t)}_{\pseudo\CS_\text{imp}}}
  { \expval{ e^{-\beta\lambda Q}}_{\pseudo\CS_\text{imp}}}.
  \end{equation}
  Grand-canonical pseudoparticle propagators on the contour $\CC$
  are defined as 
  \begin{equation}
  \label{gpseudol}
  \pG^\lambda_{mm'}(t,t')  = -i\expval{\TC a_m(t) a_{m'}^\dagger(t')}_\lambda.
  \end{equation}
  (In the following, propagators are considered to be matrices in their flavor 
  indices $m,m'$, and the indices will be omitted whenever this is not ambiguous.) 
  
  The restricted trace in Eq.~(\ref{trace1}) can be recovered from grand-canonical 
  expectation values by means of an expansion in powers of the fugacity 
  $\zeta=e^{-\beta\lambda}$.\cite{Coleman1984a} 
  For observables which annihilate the $Q=0$ state [such as the impurity 
  Green's function (\ref{impg})], an expansion of Eq.~(\ref{gge}) in $\zeta$ 
  yields 
  \begin{align}
  \label{expvalq1}
   \frac{\expval{ \pseudo  A(t)}_\lambda}{\expval{Q}_\lambda}
   &=
   \expval{ \pseudo  A(t)}_{Q=1} + \mathcal{O}(\zeta).
  \end{align}
  Furthermore, the leading terms of the Green's functions (\ref{gpseudol}) in the fugacity 
  expansion can be obtained in the form [cf.~Eq.~(\ref{interactionrepresentation})]
  \begin{equation}
  \label{laminf}
  \pG^\lambda(t,t') = k_\lambda(t,t')[ \pG(t,t') +\mathcal{O}(\zeta)],
  \end{equation}
  where the {\em projected Green's function} $ \pG(t,t')$ is independent of
  $\lambda$, and the pre-factor is given by
  \begin{equation}
  \label{klambda}
  k_\lambda(t,t') = e^{\lambda(\text{Im}t-\text{Im}t')}
  [\Theta_\CC(t,t')+\Theta_\CC(t',t)\,\zeta\,].
  \end{equation}
  The step function $\Theta_\CC(t,t')$ is $0$ if $t$ is earlier on $\CC$ than $t'$, 
  and $1$ otherwise. In order to obtain a perturbation expansion directly in 
  terms of projected quantities, the limit $\lambda\to\infty$ must be taken 
  analytically in all expressions below. As a consequence, projected Green's functions 
  become the basic objects in the hybridization expansion.

  In addition to the symmetries of the grand-canonical propagators,\cite{keldyshintro} 
  projected propagators have a number of useful properties, which we list in the 
  following paragraph. First, one can show from their 
  definition that they satisfy an initial condition 
  \begin{equation}
  \label{pseudoinitial}
  \pG_{mm'}(t^+,t) = -i \delta_{mm'},
  \end{equation}
  when $t^+$ is infinitesimally later on $\CC$ than $t$. 
  Furthermore, the factor $k_\lambda$ essentially restricts propagation of pseudoparticles 
  to one direction along the contour. In particular, 
  the leading order of the product of two Green's functions $A^\lambda$ and 
  $B^\lambda$ is given by
  \begin{equation}
  \label{pseudoproduct}
  A^\lambda(t,t_1)B^\lambda(t_1,t') \sim k_\lambda(t,t') A(t,t_1)B(t_1,t')
  \end{equation}
  if the time arguments $t'$, $t_1$, and $t$ are in cyclic order with respect
  to the arrow in Fig.~\ref{fig-contour}, and smaller by a factor 
  $\zeta$ otherwise. [In the following, we will use the notation
  $t_1 \cyles t_2 \cyles \ldots \cyles t_n$ to indicate that time arguments 
  $t_1 \ldots t_n$ are in cyclic order along $\CC$, according to the arrow in 
  Fig.~\ref{fig-contour}.] Consequently, to leading order in $\zeta$ the contour 
  convolution of the two functions is given by
  \begin{equation}
  \label{pseudoconvolution}
  \intC 
  \!d\tb\,  A^\lambda(t,\tb)  B^\lambda(\tb,t') 
  =
  k_\lambda(t,t')
  \!\!\!\!\!
  \int\limits_{\CC,\,t' \cyles \tb \cyles t}^{} 
  \!\!\!\!\!
  \!d\tb\,  A(t,\tb)  B(\tb,t') 
  \end{equation}
  where the integral range on the right hand side must be 
  restricted such that $t' \cyles \tb \cyles t$.

  \subsection{Pseudoparticle Dyson equation}
  \label{sec-pseudodyson}
  \renewcommand{\YY}{G}
  \renewcommand{\KK}{\Sigma}
  \renewcommand{\tmin}{0}  

  Grand-canonical pseudoparticle propagators obey the usual Dyson equation
  with the pseudoparticle self-energy $\Sigma^\lambda$
  \begin{equation}
  \pG^\lambda  = \pG_0^\lambda
  +
  \pG_0^\lambda
  \convz
  \Sigma^\lambda
  \convz
  \pG^\lambda,
  \end{equation}
  where $[a \convz b](t,t')$ $=$ $\intC d\tb \,a(t,\tb)\,b(\tb,t')$ 
  denotes the contour convolution,  and 
  \begin{equation}
  \label{gpseudo0}
  \pG^\lambda_{0,mm'}(t,t') = -i \expval{a_m(t)
  a^\dagger_{m'}(t')}_{\pseudo \CS_\text{loc},\lambda}
  \end{equation} 
  is the bare pseudoparticle propagator (i.e., at zero hybridization).
  The latter satisfies the equation of motion
  \begin{align}
  \label{eom-g0}
  [i\partial_t - \lambda(t) -E(t)]
  \,
  \pG^\lambda_{0}(t,t')
  =
   \delC(t,t'),
  \end{align}
  where $\lambda(t)=\lambda$ on the imaginary  part of the contour
  and zero otherwise, and $[E(t)]_{mm'}= \delta_{mm'}E_m(t)$ is a diagonal
  matrix in the flavor indices. 
  We use the notation of Ref.~\onlinecite{Eckstein2010a} for
  the derivative $\partial_t$ and the contour delta-function
  $\delC$, i.e., the latter is defined such that 
  $\intC \!d\tb \,\delC(t,\tb) f(\tb)=f(t)$ holds for any function
  $f(t)$ on the $\CC$, and $\partial_t \Theta_\CC(t,t')=\delC(t,t')$.
  
  Although we will not need an explicit expression for the bare projected
  propagator $\pG_{0}(t,t')$ in the following, it may be a useful illustration
  to compute it from the equation of motion (\ref{eom-g0}) and verify that it 
  satisfies all the usual symmetries of the contour Green's 
  functions\cite{keldyshintro} and the initial condition (\ref{pseudoinitial}). 
  By integrating 
  Eq.~(\ref{eom-g0}) with a periodic or antiperiodic boundary condition for 
  Bose and Fermi particles, respectively, we obtain the grand-canonical 
  propagator
  \begin{align}
  \label{g0lambda-explicit}
  \pG^\lambda_0(t,t')
  &=
  -i\,e^{\lambda(\text{Im}t-\text{Im}t')}\,
  \frac{\exp[-i \int_{t'}^t d\tb E(\tb)]}{e^{\beta[\lambda+E(0)]}-\chi}
  \nonumber\\
  &\times[e^{\beta[\lambda+E(0)]}\Theta_\CC(t,t') + \chi \Theta_\CC(t',t)],
  \end{align}
  where $\chi=+1$ $(-1)$ for Bose (Fermi) particles, and $E(0)$ is the value
  on the imaginary time axis. Taking the limit $\lambda\to\infty$ in this 
  expressions yields $\pG^\lambda_0(t,t')=k_\lambda(t,t') \pG_0(t,t')$, with
  the projected propagator
  \begin{align}
  \label{g0-explicit}
  \pG_0(t,t')
  %&
  =
  -ie^{-i \int_{t'}^t d\tb E(\tb)}
  %\nonumber\\
  %&
  %\times
  [\Theta_\CC(t,t') \!+\! \chi e^{-\beta E(0)} \Theta_\CC(t',t)].
  \end{align}
  
  Using Eq.~(\ref{eom-g0}), the Dyson equation can be written in differential form
  \begin{equation}
  \label{pseudodyson-gc}
  \big[i\partial_t - \lambda(t) -E(t)\big]  \pG^\lambda(t,t') 
  - [\Sigma^\lambda \convz \pG^\lambda] (t,t') =  \delC(t,t').
  \end{equation}
  The corresponding Dyson equation for the projected propagators is then derived 
  by inserting Eqs.~(\ref{laminf}), (\ref{klambda}), and (\ref{pseudoconvolution})
  into (\ref{pseudodyson-gc}), and taking the limit $\lambda \to \infty$,
  \begin{align}
  \label{pseudodyson}
  [i\partial_t  -E(t)] \, \pG(t,t') 
  - 
  \!\!\!\!\!
  \int\limits_{\CC,\,t' \cyles \tb \cyles t}^{} 
  \!\!\!\!\! 
%  \int\limits_{\CC,t'}^{t} 
  \!d\tb\,  \Sigma(t,\tb)  \pG(\tb,t')=0.
  \end{align}
  The delta-function on the right hand side has been omitted, because
  this equation will be considered only for $t \neq t'$.   
  
  The numerical solution of Eq.~(\ref{pseudodyson}) can be performed 
  in the same way as the solution of Dyson-like equations for 
  real-particle propagators, which is described in detail in 
  Ref.~\onlinecite{Eckstein2010a}. However, the structure of the integral
  in Eq.~(\ref{pseudodyson}) implies an important simplification. In 
  Eq.~(\ref{pseudodyson}), the derivative $\partial_t \pG(t,t')$ is determined 
  entirely by the value of $\pG(t_1,t')$ for $t'\cyles t_1 \cyles t$. For fixed 
  $t'$, Eq.~(\ref{pseudodyson}) is thus a Volterra integrodifferential equation,%
  \cite{Brunner1986a} whose numerical solution is similar to that of 
  an ordinary differential equation with initial condition (\ref{pseudoinitial}). 
  This is particularly interesting for the initial state, i.e.,  when all time 
  arguments are on the imaginary branch of the contour. Substituting the 
  definition (\ref{matsubara-def}) into Eqs.~(\ref{pseudodyson-gc}) and 
  (\ref{pseudodyson}) yields
  \begin{align}
  \label{pseudodyson-mat-gc}
  (-\partial_\tau - \lambda - E)&\, \pG^{\lambda,\mat}(\tau)  
  \nonumber\\
 % &\,\,\,\,-
  &-
  \int_0^\beta\!\!\! d\bar\tau \,
  \Sigma^{\lambda,\mat}(\tau\!-\!\bar\tau)
  \pG^{\lambda,\mat}(\bar\tau) = \delta(\tau),
  \end{align}
  for the grand-canonical version of the Dyson equation,
  and 
  \begin{align}
  \label{pseudodyson-mat}
  (-\partial_\tau - E) \pG^{\mat}(\tau) -
  \int_0^\tau\! d\bar\tau \,
  \Sigma^{\mat}(\tau-\bar\tau)
  \pG^{\mat}(\bar\tau) = 0,
  \end{align}
  for the projected Dyson equation.
  While Eq.~(\ref{pseudodyson-mat-gc}) is a boundary value
  problem and must be solved by Fourier transformation 
  [$\pG^{\lambda,\mat}(\beta)=\pm \pG^{\lambda,\mat}(0^-)$ for
  bosons or fermions], the projected Eq.~(\ref{pseudodyson-mat})
  is an initial value problem [$\pG^\mat(0)=-1$], which is most 
  efficiently solved in the imaginary time domain.

  \subsection{Diagram rules for the pseudoparticle self-energy}
  \label{sec-nca-sigma}
  
\begin{figure}
   % \centerline{\includegraphics[width=0.95\columnwidth,clip=true]{Figures/fig/sigma.eps}}
    \centerline{\includegraphics[width=0.95\columnwidth,clip=true]{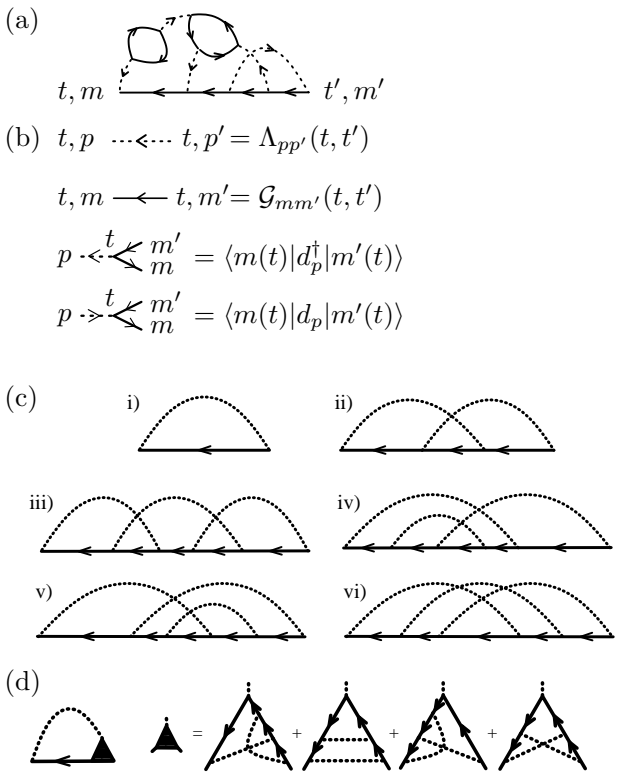}}
    \caption{
    (a) A 5th-order contribution to the self-energy $\Sigma^\lambda_{mm'}(t,t')$,
    consisting of Green's functions $\pG_0^\lambda$ (solid lines) and hybridization 
    functions $\Lambda$ (dotted lines). 
    (b) Building blocks of the diagrams in the hybridization expansion. The 
    pseudoparticle line (solid line) corresponds to the interacting propagator 
    $\pG$ in skeleton expansions, and to $\pG_0$ otherwise.
    (c) All diagrams for $\Sigma^\scell[\pG,\Lambda]$ up to third order. In the topologies  
    indicated, the hybridization line can point in any direction, which gives 
    $2$, $4$, and $4\times  8$ diagrams in first, second, and third order, respectively.
    (d) Factorization of third order diagrams [(iii)-(vi) in (c)] by separating out the 
    vertex part.
    }
    \label{fig-nca-sigma}
  \end{figure}

  Because the local part of the pseudoparticle action (\ref{pseudoaction}) 
  is quadratic, a diagrammatic expansion of pseudoparticle Green's functions 
  and self-energies in terms of $\Lambda$ can be derived from the standard 
  rules for general quartic interaction terms (see, e.g., Ref.~\onlinecite{Negele1988a}). 
  Each diagram for $\Sigma^\lambda$ contains one sequence of pseudoparticle 
  propagators that connect the two external vertices (the ``backbone''), 
  and possibly additional loops of propagator lines, e.g., renormalizations 
  of the hybridization function (Fig.~\ref{fig-nca-sigma}a). To leading order 
  in $\zeta$, the backbone $\pG^\lambda(t,t_n) \cdots \pG^\lambda(t_2,t_1)\pG^\lambda(t_1,t')$ 
  is given by $k_\lambda(t,t') \pG(t,t_n) \cdots \pG(t_2,t_1) \pG(t_1,t')$ if $t_1,\ldots,t_n$ 
  are in cyclic order along $\CC$, and smaller by $\mathcal{O}(\zeta)$ if the 
  vertices are not ordered [cf.~Eq.~(\ref{pseudoproduct})]. Each closed loop 
  of pseudoparticle propagators contributes an additional exponentially small 
  factor $\zeta$. Thus the diagram rules for the projected self-energy 
  $\Sigma(t,t') = \Sigma^\lambda(t,t')/k_\lambda(t,t')$ can be obtained 
  from the diagram rules for $\Sigma^\lambda$ by (i), replacing pseudoparticle 
  propagators (\ref{gpseudol}) by projected propagators (\ref{laminf}), (ii), 
  discarding diagrams with closed loops, and (iii), requiring vertices 
  along the backbone to be in cyclic order along $\CC$.

  For completeness we summarize the final rules for constructing the 
  projected  self-energy $ \Sigma(t,t')$: 
  (i) The $n$th order contribution to $ \Sigma(t,t')$ is given 
  by all diagrams consisting of $2n$ three-leg vertices 
  (Fig.~\ref{fig-nca-sigma}b)
  at times $t_0=t'$,$t_1$,...,$t_{2n}=t$, of which $n$ correspond to 
  annihilation operators $d$ (outgoing hybridization line), and $n$ 
  correspond to $d^\dagger$ (ingoing hybridization line). 
  The vertices are labeled according to Fig.~\ref{fig-nca-sigma}b.
  They are connected by one sequence of pseudoparticle lines (solid lines,
  pointing from $t'$ to $t$), and $n$ hybridization lines (dotted lines)
  in all possible ways such that the diagram cannot be separated into two parts by cutting only one line.  
  (ii) Sum over all internal flavor indices, and integrate over the internal 
  times $t_1$,...,$t_{2n-1}$, respecting the cyclic order 
  $t'$ $\cyles$ $t_1$ $\cyles ... \cyles$ $t$. (iii) 
  Because exactly one fermionic pseudoparticle operator is attached
  to each end of a hybridization line, the sign of the diagram is 
  $(-1)^{s+f}$, where $s$ is the number of crossing of hybridization lines, 
  and $f$ is the number hybridization lines that point opposite to the 
  direction of the backbone. (iv) An overall factor $i^n$ msut be 
  added.
  
  The diagrammatic expansion for $\Sigma$ can be resummed by replacing bare propagators 
  with interacting propagators $\pG$, and in turn taking into account only skeleton diagrams, 
  i.e., diagrams in which internal propagator lines have no self-energy insertions. Truncation 
  of the skeleton series $\Sigma^\scell[\pG,\Lambda]$ at finite order leads to conserving 
  approximations,\cite{Baym1961a} because it can be derived from the Luttinger-Ward 
  functional.\cite{Luttinger1960a} In particular, this fact ensures the conservation 
  of the pseudoparticle number (\ref{q}), which is crucial in order to obtain a meaningful 
  approximation scheme for nonequilibrium situations. To leading order in $\zeta$, the 
  conservation of $\expval{Q}_\lambda$ implies
  \begin{align}
  \label{pseudo-q}
  \tilde Q
  \equiv\lim_{\lambda\to\infty} 
  \frac{\expval{Q}_\lambda}{\zeta}
  &=
  i\sum_m (-1)^m \pG_m(t,t^+) 
  \\
  \label{pseudo-q1}
  &=
  -\sum_m \pG_m^\mat(\beta),
  \end{align}  
  where $t^+$ is infinitesimally later on $\CC$ than $t$, and $(-1)^m=\pm 1$ if $m$ 
  corresponds to Bose or Fermi particles, respectively. These relations provide 
  a good check for the numerical implementation.

  All skeleton diagrams up to third order are displayed in Fig.~\ref{fig-nca-sigma}c. The 
  self-consistent strong-coupling expansion has been proposed long ago%
  \cite{Keiter1971a,Grewe1981a,Kuramoto1983a}as an approximate solution for the Anderson 
  impurity model. Kuramoto\cite{Kuramoto1983a} coined the term non-crossing approximation 
  for the lowest order, i.e., keeping only the first diagram in Fig.~\ref{fig-nca-sigma}c. 
  In the present work we use the skeleton series up to third order as an impurity solver 
  within nonequilibrium DMFT, and compare the results to CTQMC (Sec.~\ref{sec-qmccomparison}).

  \subsection{Diagram rules for the impurity Green's function}
  
  In general, expressions for observables in the impurity model can be derived from the grand 
  potential $\Omega_\lambda = -\beta^{-1} \log \TR[ \zeta^Q \TC e^{\mathcal{\tilde S}_\text{imp}}]$.
  Because diagrams for the correction $\Delta\Omega_\lambda  = \Omega_\lambda-\Omega_\lambda(\Lambda=0)$ 
  contain at least one closed loop of pseudoparticle lines, $\Delta \Omega_\lambda$ is proportional 
  to $\zeta$ for $\lambda\to\infty$. The leading order in $\zeta$,
  \begin{equation}
  \label{pseudoomega}
  \Omega = \lim_{\lambda\to\infty} \frac{-1}{\zeta\beta}
  \log \TR[\zeta^Q \TC e^{\mathcal{\tilde S}_\text{imp}}]
  \end{equation}
  is obtained by adding to the local contribution $\Omega(\Lambda=0)$ all diagrams of 
  $\Omega_\lambda$ which contain only one loop, in which $\pG^\lambda$ is replaced by 
  $\pG$, and where integrals over the internal vertices are restricted such that the 
  vertices are in cyclic order. (See the analogous argument for $\Sigma$ in 
  Sec.~\ref{sec-nca-sigma}.) 
  
\begin{figure}
    %\centerline{\includegraphics[width=0.8\columnwidth]{Figures/fig/nca-g.eps}}
    \centerline{\includegraphics[width=0.8\columnwidth]{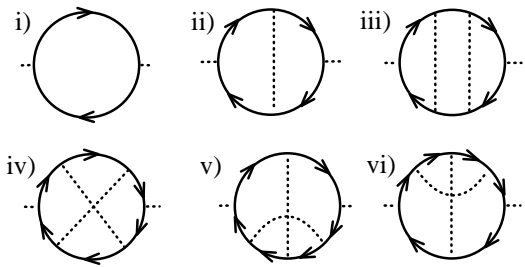}}
    \caption{All diagrams for $G^\scell[\pG,\Lambda]$ in first order [diagram (i)],
    second order [diagram (ii)], and third order [diagram (iii)-(vi)]. Internal 
    hybridization lines can point in any direction, which gives $1$, $2$, and $4\times  4$ 
    diagrams in first, second, and third order, respectively.}
    \label{fig-nca-g}
  \end{figure}

  Using Eq.~(\ref{expvalq1}), the impurity Green's function (\ref{impg}) is given 
  by $G(t,t') = \lim_{\lambda\to\infty}  G^\lambda(t,t')/\expval{Q}_\lambda$,
  where $G_{pp'}^\lambda(t,t')= -i \expval{\TC \tilde d_p(t) \tilde d_{p'}^\dagger(t')}_\lambda$.
  It can thus be obtained from the derivative [cf.~Eqs.~(\ref{pseudo-d}), (\ref{pseudoaction}), 
  (\ref{pseudo-q}), and (\ref{pseudoomega})]
  \begin{equation}
  \label{gd1}
  G_{pp'}(t,t') 
  =
  \frac{\beta}{\tilde Q}\frac{\delta \Omega }{\delta \Lambda_{p'p}(t',t)}.
  \end{equation}
  Diagrams for $G(t,t')$ (in terms of the projected pseudoparticle 
  Green's functions) are therefore constructed by removing one hybridization line 
  from the diagrams for $\Omega$ (Fig.~\ref{fig-nca-g}). Note that a diagram for $\Omega$ generally has 
  a symmetry factor $1/S\neq 1$, where $S$ is the number of topologically equivalent 
  ways to label the vertices. The symmetry factor disappears in the expansion of 
  $G$, because for a diagram with symmetry factor $1/S$ there are $S$ ways to 
  remove a hybridization line which lead to the same diagram for $G$. The 
  series for $G$ can thus be resummed in the same way as the series for $\Sigma$, 
  i.e., by keeping only skeleton diagrams for $G$, and replacing $\pG_0$ with $\pG$. 
  Equation (\ref{gd1}) then holds also for the skeleton expansion,
  \begin{equation}
  \label{gdscell}
  G^\scell_{pp'}[\pG,\Lambda](t,t') = \frac{\beta}{\tilde Q} \frac{\delta \Omega^\scell[\pG,\Lambda] }
  {\delta \Lambda_{p'p}(t',t)},
  \end{equation}
  where $\Omega^\scell[G,\Lambda]$ is the Luttinger Ward functional, i.e., the 
  skeleton expansion for $\Omega$ in terms of the fully interacting (projected) 
  propagators $\pG$. To design an approximation for $G^\scell[\pG,\Lambda]$ which 
  is consistent with a given approximation of $\Sigma$ one must truncate both 
  $\Omega^\scell[\pG,\Lambda]$ and $\Sigma^\scell[\pG,\Lambda]$ at the same order.

  The final rules for $G^\scell[\pG,\Lambda]$ read:
  The $n$th order contribution consists of a loop of projected pseudoparticle 
  propagator lines (Fig.~\ref{fig-nca-sigma}b) which connects $2n$ vertices ($n$ 
  annihilation operators, $n$ creation operators). One $d$-vertex (time $t$, 
  $\Lambda$-line labeled $p$) and one $d^\dagger$-vertex (time $t'$, $\Lambda$-line 
  labeled $p'$) are external vertices. The internal vertices are connected 
  by hybridization lines such that no internal line has a self-energy insertion.
  Sum over all internal flavor indices, and integrate over internal (contour) 
  time variables respecting the cyclic order of $t_1 \ldots t_{2n}$ along 
  the contour. Add a pre-factor $i^n$. To determine the sign of a diagram 
  $D$, reinsert the $\Lambda$-line between the external vertices. This recovers 
  the diagram $D'$ in the expansion of $\Omega$ from  which the diagram $D$ 
  is derived. The sign $\kappa_D$ of $D$ is given by the sign of $D'$, i.e., 
  $\kappa_D=(-1)^{s+f}$, where $s$ is the number of crossings of hybridization 
  lines, and $f$ is the 
  number of hybridization lines that go in opposite direction to the string 
  of pseudoparticle lines that is obtained if the loop is cut at an arbitrary
  fermionic propagator line. All skeleton diagrams for $G$ up to third order are
  shown in Fig.~\ref{fig-nca-g}.
  
  \subsection{Numerical implementation}
   
  \label{sec-implementation}

  Before presenting first benchmark results for the self-consistent hybridization
  expansion, we would like to make some remarks on the numerical implementation.
  First of all, we note that the number of possible labelings for the internal
  flavor indices is usually quite restricted. As an example, consider the 
  single-impurity Anderson model with the four basis states $\ket{0}$, 
  $\ket{\sigma}=d_\sigma^\dagger\ket{0}$, 
  and $\ket{\uparrow\downarrow}=d_\downarrow^\dagger d_\uparrow^\dagger\ket{0}$,
  $H_\text{loc}= U d^\dagger_\uparrow d_\uparrow d^\dagger_\downarrow d_\downarrow
  -\mu(d^\dagger_\uparrow d_\uparrow +d^\dagger_\downarrow d_\downarrow)$, and 
  $S_{hyb}=-i\intC dt dt' \sum_\sigma d^\dagger_\sigma(t) \Lambda_\sigma(t,t') 
  d_\sigma(t')$. The matrix elements (\ref{d-matrixelement}) are nonzero only for 
  the combinations $F_{\sigma,0}^\sigma=1$ and $F_{\uparrow\downarrow,\sigma}^{\bar\sigma}=\sigma$.
  Furthermore, Green's functions are diagonal in pseudoparticle flavor, because both the
  interaction part and the hybridization function are diagonal in the occupation
  number basis. The second-order digram for $\Sigma_{mm}(t,t')$,
  e.g., then only allows eight possible labelings for the three internal Green's function 
  lines, which are ($\uparrow,0,\downarrow$) and ($\downarrow,0,\uparrow$) 
  for $m=\uparrow\downarrow$, ($0,\bar\sigma,\uparrow\downarrow$) and 
  ($\uparrow\downarrow,\bar\sigma,0$) for $m=\sigma$, and 
  ($\uparrow,\uparrow\downarrow,\downarrow$) and ($\downarrow,\uparrow\downarrow,\uparrow$) 
  for $m=0$.
  
  To obtain a self-consistent solution, the hybridization expansion is evaluated by iteratively
  solving the Dyson equation (\ref{pseudodyson}) for $\pG(t,t')$, and evaluating the integral 
  expressions for $\Sigma(t,t')$. However, the real-time version of the expansion can easily 
  be implemented in a slightly simpler way that exploits the causal structure of the equations.
  If we have computed $\pG(t,t')$ for $\text{Re}(t)$, $\text{Re}(t')$ $\le \tmax$, then 
  $\pG(\tmax+\Delta t,t')$ and $\pG(t,\tmax+\Delta)$ can be obtained from the above mentioned 
  iteration in only one or two steps by starting from a polynomial extrapolation of $\pG(t,t')$. 
  This amounts to a stepwise propagation of the solution on the imaginary branch of $\CC$ to real 
  times.
  
  The numerical effort of the evaluation of diagrams is mainly determined by the 
  contour integrals over the internal vertices. Using Monte Carlo for the evaluation 
  of the integrals for higher order diagrams will suffer from a sign problem. 
  We use a quadrature formula for the integrals which is based on equidistant 
  discretization of the contour $\CC$. The $n$th order diagrams have $2n-2$ internal 
  integrals  
  (cf.~Figs.~\ref{fig-nca-sigma}c 
  and \ref{fig-nca-g}), which have to be evaluated for each combination of the two 
  external time variables. (Nonequilibrium correlation functions depend on both time 
  arguments separately). This seems to imply that the numerical effort for the evaluation 
  of $\Sigma$ and $G$ scales with the number $N$ of mesh points like $N^2$, $N^4$, and 
  $N^6$ for first, second and third order, respectively. (The effort for the solution 
  of the Dyson equation, which is essentially a matrix inversion on $\CC$, scales 
  like $N^3$.) However, one can reduce the effort for the evaluation of the third-order 
  diagrams to $N^5$ by factorizing out a vertex part $\tilde F$ with two internal 
  integrals and three external variables (Fig.~\ref{fig-nca-sigma}d); $\Sigma$ and $G$ 
  can then be computed from $\tilde F$ with only two additional integrals. Since 
  $\tilde F$ does not have to be stored in memory, this way of evaluation is more 
  efficient than performing four internal integrals.

  %\subsection{Determination of the impurity self-energy}
  %secret
    
  \section{Comparison to CTQMC}
  \label{sec-qmccomparison}
  
   \subsection{Interaction quench in the Hubbard model}
   
   Nonequilibrium DMFT for the interaction quench in the Hubbard model provides 
   a perfect framework to benchmark the perturbative impurity solver. The 
   Hubbard Hamiltonian
   \begin{align}
    \label{hubbard}
    H(t)
    &=
    \sum_{ij\sigma}
    V_{ij}
    c_{i\sigma}^{\mydagger}
    c_{j\sigma}^{\phdagger}
    +
    U(t)
    \sum_{i}
    \big(n_{i\uparrow}\!-\!\tfrac12\big)
    \big(n_{i\downarrow}\!-\!\tfrac12\big)
   \end{align}
   describes fermions of spin one half which hop on a lattice
   with hopping amplitude $V_{ij}$ and interact with a repulsion energy 
   $U$ on each site. To perform an interaction quench, the system is 
   prepared in a thermal equilibrium state at temperature $T=1/\beta$
   and interaction $U(t<0)=U_0$ for times $t<0$, and the interaction 
   is suddenly switched to a new value $U(t>0) = U$ at  $t=0$. 

   The DMFT equations for the interaction quench have been explained in
   detail in Refs.~\onlinecite{Eckstein2009a} and \onlinecite{Eckstein2010a}. 
   In the following we assume that the hopping matrix $V_{ij}$ has a semielliptic 
   local density of states 
   \begin{equation}
   \label{bethedos}
   \rho(\epsilon)=\sqrt{4-(\epsilon/V)^2}\,/\,2\pi V
   \end{equation}
   (with half-bandwith $2V$), and we focus on the paramagnetic state at 
   half-filling. DMFT then reduces to a set of two self-consistent 
   equations:\cite{Eckstein2009a,Eckstein2010a} (i) The local lattice Green's function (\ref{impg}) 
   must be determined from the single-site action (\ref{action}) , 
   where the index $p$ now labels spin $\sigma=\uparrow,\downarrow$, 
   and the local Hamiltonian is given by
   \begin{equation}
    H_\text{loc}(t)=
    U(t)
    \sum_{i}
    \big(n_{i\uparrow}\!-\!\tfrac12\big)
    \big(n_{i\downarrow}\!-\!\tfrac12\big).
   \end{equation}
   (ii) The hybridization function is determined by
   the self-consistency\cite{Eckstein2010b}
   \begin{equation}
   \label{sce}
   \Lambda_\sigma(t,t') = V^2 G_\sigma(t,t').
   \end{equation}
   The hopping $V=1$ is used as an energy unit, and times are measured
   in units of the inverse hopping ($\hbar=1$).
   
   Below we solve these DMFT equations by means of the self-consistent hybridization 
   expansion and compare to results from CTQMC.\cite{Eckstein2009a,Eckstein2010a} 
   In particular, we focus on the time-evolution and the thermal equilibrium value 
   of the double occupancy per site, $d(t)=\expval{n_{i\uparrow}n_{i\downarrow}}$, 
   which is a local observable  and can thus be measured directly in the impurity 
   model, i.e.,
   \begin{equation}
   d(t)=i \,\tilde Q^{-1}\,\pG_{\uparrow\downarrow}(t,t^+),
   \end{equation}
   where $\pG_{\uparrow\downarrow}$ is the propagator for doubly occupied sites,
   and $t^+$ is infinitesimally later on $\CC$ than $t$.

   \subsection{The initial equilibrium state}

   \begin{figure}
   %\centerline{\includegraphics[width=1\columnwidth]{Figures/plots_neu/dmft_bethe_docc_eq.eps}}
   \centerline{\includegraphics[width=1\columnwidth]{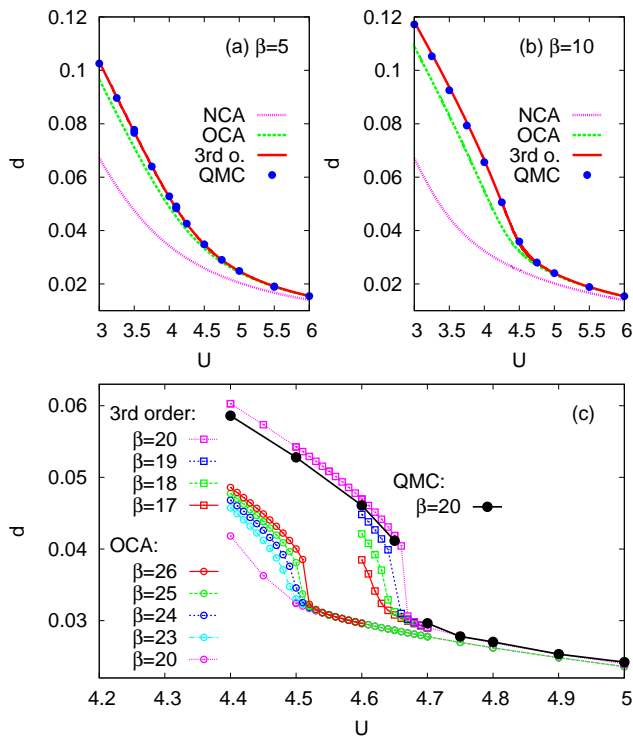}}
   \caption{Double occupancy $d_{eq}(\beta,U)$ in the thermal equilibrium state. The 
   impurity solver is either CTQMC, or the self-consistent hybridization expansion 
   up to first (NCA), second (OCA), and third order. (a) $\beta=5$. (b) $\beta=10$. 
   (c) $\beta$ close to the critical temperature $T_c$ of the first-order metal-insulator 
   transition.}
   \label{fig-docc-eq}
   \end{figure}  
   
   Figure \ref{fig-docc-eq} shows the double-occupancy $d_{eq}(\beta,U)$ in the 
   thermal equilibrium state at interaction $U$ and inverse temperature $\beta$. 
   At large enough temperature, $d_{eq}(T,U)$ decreases smoothly as a function 
   of $U$ (Fig.~\ref{fig-docc-eq}a). As $T$ is lowered, the curves bend 
   strongly around $U=4.5$ (Fig.~\ref{fig-docc-eq}b), indicating a narrow 
   crossover between metallic and insulating behavior. Below a critical 
   temperature $T_c$, the transition between metal and insulator becomes 
   a first-order phase transition (Fig.~\ref{fig-docc-eq}c). For the 
   semielliptic density of states, the endpoint of this Mott transition 
   is located at $U_c=4.7$ and $T_c=0.055$.\cite{Bluemer-thesis}  
   
   In agreement with recent results based on a Monte Carlo sampling around 
   NCA,\cite{Gull2010a} we find that NCA can reproduce the CTQMC results 
   only deep in the insulating phase. However, already the lowest order 
   correction to NCA, i.e., OCA, very well accounts for the nonlinear 
   behavior of $d_{eq}(U,T)$ in the crossover regime, and the third order 
   in the self-consistent hybridization expansion almost quantitatively 
   recovers the CTQMC results even close to the critical point 
   (Fig.~\ref{fig-docc-eq}c). The location of the critical 
   endpoint in the phase diagram is in good agreement with previous QMC 
   results:\cite{Bluemer-thesis} If we estimate $T_c$ 
   from the smallest $\beta$ for which we can detect hysteretic behavior in 
   $d_{eq}(\beta,U)$ (this gives actually a lower bound for $T_c$), we find 
   $T_c > 1/19 \approx 0.052$ for the 3rd order and $T_c > 1/26 \approx 0.038$ 
   for OCA (Fig.~\ref{fig-docc-eq}c). NCA, on the other hand, does not display 
   singular behavior in this parameter regime. As usual, the convergence of 
   the DMFT equations slows down close to the critical point, and it is thus 
   hard to get precise numbers for $T_c$ and $U_c$.
   
\begin{figure}
   %\centerline{\includegraphics[width=1\columnwidth]{Figures/figure03.eps}}
   \centerline{\includegraphics[width=1\columnwidth]{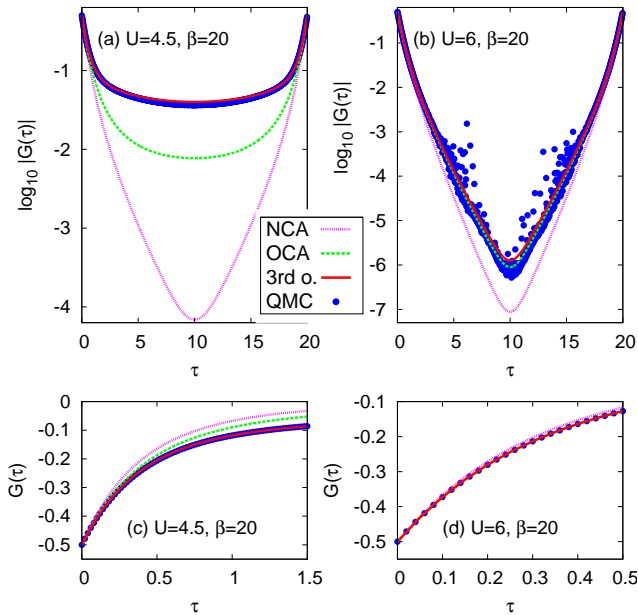}}
   \caption{
   Matsubara component (\ref{matsubara-def}) of the local Green's function
   $G_\sigma(\tau)$ [Eq.~\ref{impg}], which is independent of $\sigma$ in the 
   paramagnetic phase. The impurity solver is either CTQMC, or the self-consistent 
   hybridization expansion up to first (NCA), second (OCA), and third order. (a) 
   $U=4.5, \beta=20$ (crossover regime). (b) $U=6, \beta=20$ (insulating phase). 
   Note that  the CTQMC results in the insulating phase are only accurate for
   values larger than the ``noise floor'' of about $10^{-3}$.  Panels (c) and 
   (d) show the same data as (a) and (b), respectively, but  plotted for small 
   $\tau$ on a linear scale.
   }
   \label{fig-gloc-eq}
   \end{figure}

   The order-by-order convergence of the self-consistent hybridization expansion 
   is also evident from the local Green's function $G(\tau)$, both in the crossover
   regime (Fig.~\ref{fig-gloc-eq}a) and in the insulating phase (Fig.~\ref{fig-gloc-eq}b).
   From the value $G(\beta/2)$ one can see that NCA overestimates the insulating 
   nature of the solution. This fact reflects a well known deficiency of NCA: The 
   Kondo temperature $T_K$ for the Anderson model comes out correct within NCA 
   for $U=\infty$, but it is severely underestimated for finite interaction $U$. 
   This problem can be cured by taking into account certain vertex corrections, 
   which correspond to summing up higher order terms in the self-consistent 
   expansion.\cite{Pruschke1989a,Haule2001a} Our results show that the third 
   order is sufficiently accurate in a wide parameter range covering the insulating 
   phase and the crossover regime, even close to the critical point.

   Another known deficiency of the NCA is that the Fermi-liquid in the Anderson
   impurity model for $T \ll T_K$ is not correctly described.\cite{MuellerHartmann1984a} 
   Because this problem cannot be cured by taking into account finite order diagrams 
   in the hybridization expansion,\cite{Haule2001a} one would expect that even the 
   third order will yield  wrong results in the metallic phase at low enough temperature. 
   Empirically, we find a slow-down of the convergence as the temperature is lowered 
   in the metallic phase, and we have not systematically studied the breakdown of the 
   truncated self-consistent hybridization expansion deep in the metallic phase.

  \subsection{Time evolution of the double occupancy}

\begin{figure}
   %\centerline{\includegraphics[width=1\columnwidth]{Figures/plots_neu/qmctcacomparison.eps}}
   %\centerline{\includegraphics[width=1\columnwidth]{Figures/figure04.eps}}
   \centerline{\includegraphics[width=1\columnwidth]{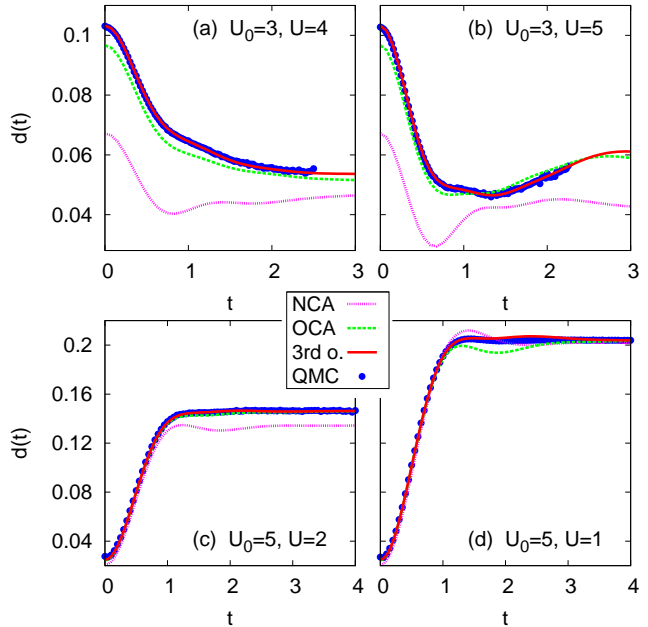}}
   \caption{Double occupancy $d(t)$ after an interaction quench from $U_0$ to $U$. 
   (a) and (b): initial states in the crossover regime [$U_0=3$, $\beta=5$], 
   (c) and (d): initial states in the insulating phase [$U_0=5$, $\beta=5$].
   }
   \label{fig-docc-neq}
   \end{figure}

  To test the accuracy of the strong-coupling expansion for nonequilibrium
  problems we compute the time-evolution of the double occupancy after an 
  interaction quench. Due to the dynamical sign problem, weak-coupling CTQMC 
  calculations 
  for interacting initial states\cite{Eckstein2010a} can be performed only 
  to relatively short times $t_\text{max}$, which mainly depend on the final 
  interaction $U(t>0)$. However, for those $t_\text{max}$ which are accessible 
  with CTQMC, the comparison with the strong-coupling expansion reveals a similar 
  trend as for thermal equilibrium states (Fig.~\ref{fig-docc-neq}): For quenches 
  from the crossover region to larger interaction both the initial state 
  and the time evolution is not correctly described within NCA, whereas OCA 
  is more reliable, and the third order calculation recovers the CTQMC results 
  almost quantitatively (Figs.~\ref{fig-docc-neq}a and b). 
  While Figs.~\ref{fig-docc-neq}a and b show the longest times accessible with 
  CTQMC, the OCA and the 3rd order calculations can be carried to substantially 
  longer times. For a quench from the insulating phase to smaller interaction, 
  NCA is better suited to describe the initial state, but differences to CTQMC 
  become more pronounced during the time evolution (Figs.~\ref{fig-docc-neq}c 
  and d). 

  \section{Modulation spectroscopy on the Mott insulating phase}
  \label{sec-spectroscopy}
  
  \subsection{Introduction}
  
  To illustrate the capability of the approach, we are now going to present an 
  application of the strong-coupling hybridization expansion in a parameter regime 
  where the weak-coupling CTQMC approach would be numerically too expensive. Our 
  aim is to compute the response of the double occupancy in the Mott insulator to 
  a time-dependent change of the interaction $U$ or the hopping amplitude $V$. Such 
  an experiment, with a periodic change of the hopping, was originally proposed by 
  Kollath {\em et al.}\cite{Kollath2006a} as a new type of spectroscopy for ultracold 
  gases in optical lattices without direct analogon in solid state physics. In the 
  meantime, the technique has been used as an experimental probe for the detection 
  of the Mott insulating phase of ultracold $\rm{^{40}K}$-atoms in an optical lattice.%
  \cite{Joerdens2008a} 

  In the experiments, the hopping amplitude $V(t)=V_0[1+\alpha \cos(\omega t)]$ is 
  modulated sinusodially over several tens of 
  periods $2\pi /\omega$.\cite{Joerdens2008a,Strohmeier2010a} Apart from an oscillating 
  component $d_{osc}(t)$ with zero time-average, the double occupancy $d(t)$ rises 
  linearly in time for small times, and saturates within a timescale $\tau_{sat}$. 
  In the Mott insulating phase, the modulation spectrum, i.e., the magnitude of the 
  response as a function of the frequency, has a peak when $\omega$ is approximately 
  at resonance with the energy $U$ that is needed for the creation of a doublon-holon 
  pair, and a gap at $\omega=0$.\cite{Kollath2006a,Huber2009a}
  
  Because the modulation strength can be quite large, many aspects of those experiments
  can only be understood by means of a nonequilibrium formalism. This is certainly the 
  case for the saturation time $\tau_{sat}$,\cite{Hassler2009a} and for the nonequilibrium 
  quasisteady (time-periodic) state which the system is in once it has saturated. Even 
  when averaged over time, such a state might have properties which do not resemble any 
  thermal equilibrium state of the system. Nonequilibrium DMFT can be used to resolve 
  some of these issues. In the following, we demonstrate that DMFT yields a modulation 
  spectrum which is in agreement with a recent investigation based on slave-boson 
  mean-field theory,\cite{Huber2009a} and similar to what has been obtained in time-dependent 
  density-matrix renormalization group calculations for the one-dimensional Hubbard model.%
  \cite{Kollath2006a} Furthermore, we will show that a slightly different modulation procedure, 
  namely a quench of the hopping or the interaction by a few per cent, provides another probe 
  which is sensitive to the Mott transition and the metal-insulator crossover at higher 
  temperatures.

  \subsection{Periodic modulation of $U$}
  
  In the following we consider the Hubbard model (\ref{hubbard}) with a time-dependent
  interaction 
  \begin{equation}
  \label{umodulation}
  U(t>0) = U_0[1+\alpha \cos(\omega t)],
  \end{equation}
  where $\alpha$ is the relative modulation strength ($\alpha < 1$). For times $t<0$, the 
  system is prepared in an equilibrium state at interaction $U(t<0)=U_0$ and temperature 
  $T=1/\beta$. The model is treated within nonequilibrium DMFT, assuming a semielliptic 
  density of states (\ref{bethedos}), such that the self-consistency is given by 
  Eq.~(\ref{sce}). The energy scale is fixed by the quarter bandwidth $V=1$. Because 
  we will restrict the investigation to insulating states and to the crossover regime, 
  we will mainly use OCA as an impurity solver.
  
  Experimentally, the hopping is more easily tunable than the interaction, because the
  former is strongly affected by a change of the optical lattice depth. Our description 
  (\ref{umodulation}) is nevertheless justified, because modulation of $U$ and $V$ are 
  equivalent from a theoretical point of view, and only the ratio $U/V$ matters. The 
  Hubbard model $H(t)$ with time-dependent interaction (\ref{umodulation}) 
  and time-independent hopping $V_0$ can be mapped onto an equivalent model with 
  time-independent interaction $U_0$ and periodic hopping 
  $V(t')=V_0/\{1+\alpha \cos[\omega t(t')]\}$, where $t(t')$ is the inverse of the 
  transformation $t'(t)=t+\alpha\sin(\omega t)/\omega$. To lowest order in $\alpha$, 
  modulation of $U(t)=U_0[1+\alpha \cos (\omega t)]$ and $V(t')=V_0[1-\alpha \cos (\omega t')]$ 
  are thus equivalent, although higher harmonic terms $\propto \alpha^n \cos(n \omega t')$ 
  appear in $V(t')$ for large $\alpha$. The equivalence can be established by 
  a simple change of time-variables in the time-evolution operator $\mathcal{U}(t,0) = 
  T_t \exp[-i\int_0^{t} d\bar t H(\bar t)]$ from $t$ to $t'$, which yields 
  $\mathcal{U}'(t',0)=\mathcal{U}(t(t'),0)$, where $ \mathcal{U}'(t',0)$
  %$=T_t \exp[-i\int_0^{t} d\bar t H(\bar t)]$ 
  is the time-evolution operator for the model $H'(t')$ with periodically 
  modulated hopping $V(t')$.

  \begin{figure}
  %\centerline{\includegraphics[width=1\columnwidth]{Figures/plots_neu/dgdo_u7.eps}}
  %\centerline{\includegraphics[width=1\columnwidth]{Figures/plots_neu/dgdo_u7_tsat.eps}}
  \centerline{\includegraphics[width=1\columnwidth]{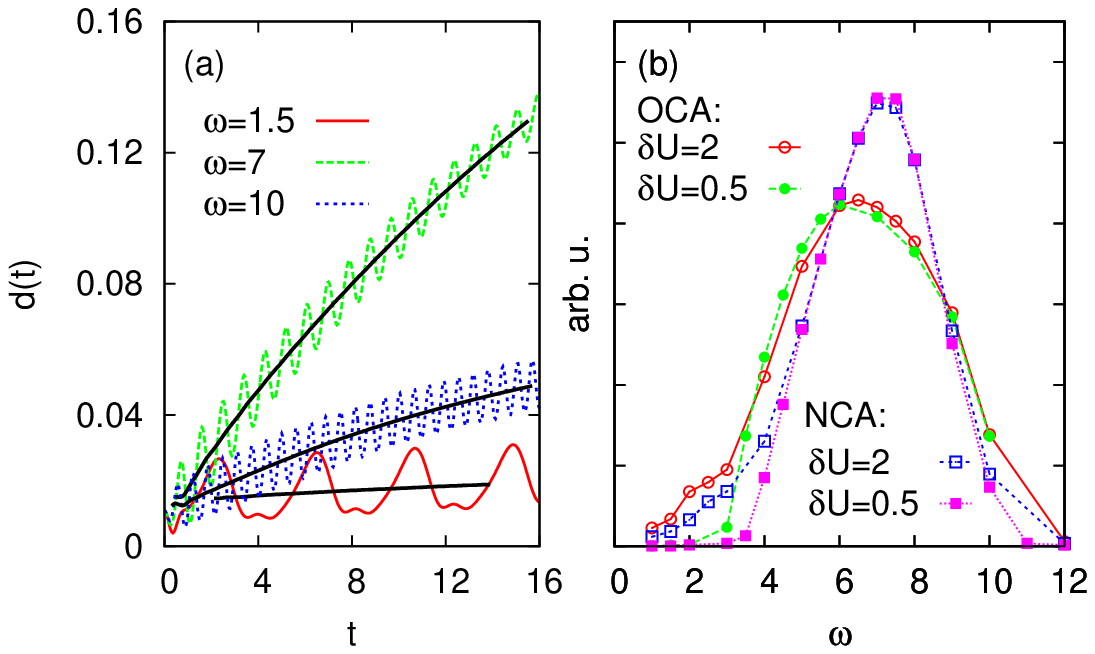}}
  \centerline{\includegraphics[width=1\columnwidth]{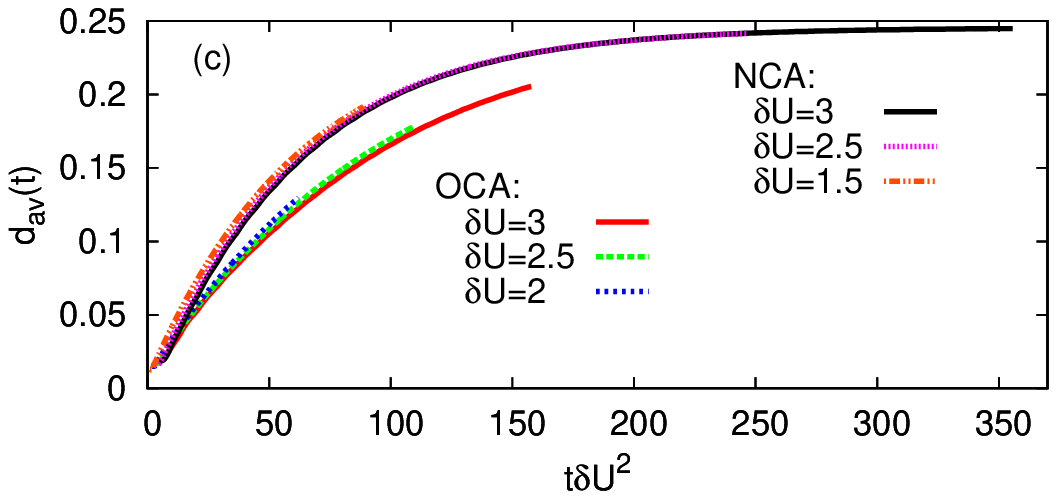}}  
  \caption{
  (a) Double occupancy $d(t)$ during a periodic modulation of the interaction around 
  $U_0=7$ ($\beta=10$, $\delta U =\alpha U_0=2$). The frequency is below the resonance 
  peak ($\omega=1.5$), close to the resonance ($\omega=7$), and above the resonance 
  ($\omega=10$). Bold solid lines indicate the average $d_{av}(t)$ over one period 
  $[t-\pi/\omega,t+\pi/\omega]$. OCA was used as an impurity solver. 
  (b) Modulation spectrum for $U_0=7$ and $\beta=10$, 
  using either NCA or OCA as an impurity solver: For large amplitude $\delta U=2$,
  the data have been obtained from $d_{av}(t_0)$ at given time $t_0=10$ (open symbols);
  for small amplitudes $\delta U=0.5$, the slope $(d/dt) d_{av}(t)$ in the interval 
  $8<t<10$ is plotted (full symbols). The curves have been scaled with a constant 
  factor to match their peak heights. (c) Averaged double occupancy $d_{av}(t)$ at 
  $U_0=7$ for various modulation amplitudes $\delta U = \alpha U_0$, plotted against 
  $t\delta U ^2$. The three curves labeled NCA almost fall on top of each other.}
  \label{fig-docct}
  \end{figure}
  
  The typical time-evolution of the double occupancy $d(t)$ during a periodic modulation
  of $U$ is displayed in Fig.~\ref{fig-docct}a. The effect of the modulation vanishes 
  for $\omega \to 0$, and is largest close to $\omega\approx U$. The anharmonic behavior
  in $d(t)$ for small frequencies $\omega$ is due to a rather large amplitude 
  $\delta U \equiv \alpha U_0$.  After removing the oscillating component $d_{osc}(t)$ 
  by taking an average  $d_{av}(t) = \int_{t-\tau/2}^{t+\tau/2} dt'\, d(t')$ over one 
  period $\tau=2\pi/\omega$, one can clearly identify an initial linear increase with 
  a slope $\Gamma(\omega)$, followed by a trend towards saturation at longer times. 

  The rate $\Gamma(\omega)$ can be obtained from second order time-dependent perturbation 
  theory, i.e., $\Gamma \propto \alpha^2$ for $\alpha\ll 1$.\cite{Kollath2006a,Huber2009a} 
  (Linear response yields only an oscillating contribution.) Since the resulting Fermi's 
  golden rule expression for $\Gamma$ depends only on the equilibrium state of the system 
  at $U=U_0$,\cite{Huber2009a} it would be best to measure $\Gamma(\omega)$ directly in 
  the limit $\alpha \to 0$. To extract the slope $\Gamma(\omega)$, one has to go to 
  small amplitudes $\alpha$ anyway, because only then does the linear region extend 
  over many oscillation periods. On the other hand, experiments are performed at quite 
  large $\alpha$ in order to obtain a good signal to noise ratio, and the magnitude of 
  the response is defined by the value of the double occupancy $d_{av}(t)$ at given time 
  $t_0$. Usually, $t_0$ is chosen so large that $d_{av}(t)$ is no longer linearly increasing 
  at $t=t_0$. 
  
  In Fig.~\ref{fig-docct}b we compare two ways to quantify the response: Either (i), 
  the increase of the double occupancy $d_{av}(t_0)-d(0)$ is measured at a given time 
  $t_0$ for large modulation amplitudes as a function of frequency, or (ii), the slope 
  $\Gamma(\omega)$ is obtained from a linear fit to the initial increase for smaller 
  amplitudes $\alpha$. Both approaches give a modulation spectrum with a peak at 
  $\omega\approx U$, and a gap at $\omega=0$. Similar to the findings of the previous 
  section, the NCA solution slightly overestimates the insulating nature of the solution 
  compared to the more reliable OCA. Like in the one-dimensional case,\cite{Kollath2006a} 
  our data show that the location of the peak at $\omega \approx U$ is not considerably 
  shifted if the measurement is no longer performed at $\alpha \ll 1$. The gap, on the
  other hand, is most reliably extracted from the second approach (ii). In the following 
  we will only focus on the peak, and not investigate the low frequency weight in detail.

\begin{figure}
  %\centerline{\includegraphics[width=1\columnwidth]{Figures/plots_neu/dgdo_max.eps}}
  \centerline{\includegraphics[width=1\columnwidth]{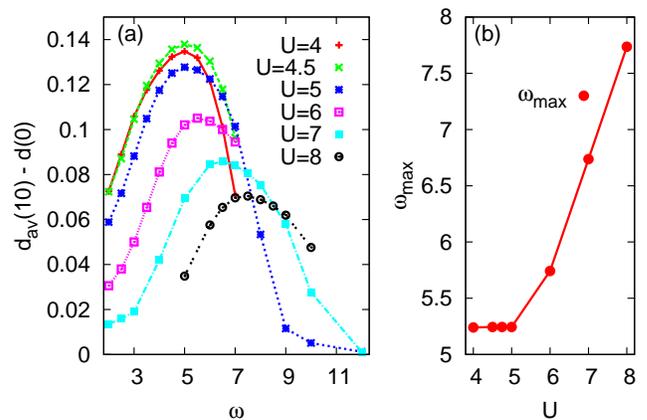}}
  \caption{
  (a) Modulation spectrum at $\beta=10$ and various interactions $U_0$, obtained 
  from the increase $d_{av}(t_0)-d(0)$ at $t_0=10$ ($\delta U=2$). (b) Location of 
  the maximum $\omega_{\text{max}}$ of the spectra in (a), plotted against $U_0$.
  }
  \label{fig-dgdo}
  \end{figure}

  Figure \ref{fig-dgdo}a displays the peak in the modulation spectrum for various values 
  of $U_0$. For the semielliptic density of states, the first-order phase transition line 
  terminates at $U_c\approx 4.7$ and $T_c\approx 0.055$,\cite{Bluemer-thesis} and the
  zero-temperature transition is located at $U_{c2}\approx 6$. Hence the data in 
  Fig.~\ref{fig-dgdo}, which are computed at $T=0.1$, correspond to a cut through the 
  crossover region of the metal-insulator phase diagram. The peak in the modulation 
  spectrum is clearly visible all throughout the insulating phase and the crossover 
  regime between metal and insulator. Its position $\omega_\text{max}(U_0)$ scales 
  linearly with $U_0$ in the insulating phase (Fig.~\ref{fig-dgdo}b), while in the 
  crossover regime we find only a weak dependence on $U_0$. This finding is in good 
  agreement with results for $\Gamma(\omega)$ from slave-boson mean-field theory,%
  \cite{Huber2009a} where peaks around $\omega\approx U$ and $\omega\approx U_{c2}$ 
  are predicted for the insulating phase and the metallic phase, respectively. In the 
  slave-boson approach, these spectral features arise from excitations between the 
  Hubbard bands in the insulator, and between pre-formed Hubbard bands in the metallic 
  phase. 
  
  Since our calculation is performed at temperatures above $T_c$ and the pre-formed Hubbard 
  bands shift with $U$, the almost kink-like $\omega_\text{max}(U_0)$ seems very remarkable. 
  It should be investigated whether the third order in the expansion leads to a smoothening 
  of this feature. This will require quite some numerical effort (although it is still feasible 
  using small-scale parallelization), and it is thus left to future work, which should 
  involve a more realistic setup.
  
  Another interesting topic is the saturation of $d_{av}(t)$ at large times.\cite{Hassler2009a}
  A scaling of the time-axis with $\alpha^2$ indicates that the saturation time $\tau_{sat}$ 
  behaves like $\tau_{sat}\propto\alpha^{-2}$ in the insulating phase for frequencies $\omega=U$ 
  (Fig.~\ref{fig-docct}c), while the saturation value $d_{av}(t\to\infty)$ does not depend 
  sensitively on $\delta U$. However, due to the long times needed to reach saturation at small 
  $\delta U$, this result has so far only been computed using NCA as an impurity solver, which 
  is reliable only deep in the insulating phase. A dependence $\tau_{sat}\propto\alpha^{-2}$ 
  would be consistent with an incoherent pumping mechanism into a doublon-holon
  continuum.\cite{Hassler2009a} 
  
  \subsection{Quench-like modulation of $U$}
  
  Modulation spectroscopy for the double occupancy is in principle not restricted 
  to the periodic modulation Eq.~(\ref{umodulation}). In particular, a small interaction 
  quench from $U(t<0)=U_0$ to $U(t>0)=U_0+\delta U$, or an equivalent quench of 
  the hopping, can be viewed as a modulation experiment as well. In order to extract a 
  frequency-dependent response signal, we define the Fourier transform
  \begin{equation}
  \label{doccft}
  \tilde d(\omega) = \text{Re} 
  %\frac{1}{\omega + i0}
  \int_0^{\tmax} \!\!\!\!dt\,e^{i\omega t} \,[d(t)-d(\tmax)],
  \end{equation}
  where $\tmax$ is the maximum time reached in the simulation, and 
  $d(t)=(1/Z)\TR[ e^{-\beta H_0} e^{iHt} d e^{-iHt}]$  is the time-dependent double 
  occupancy. Using first-order perturbation theory one obtains 
  \begin{equation}
  \label{dqu}
  \tilde d(\omega) = \omega^{-1} \,\text{Im} \tilde \chi(\omega) + \mathcal{O}[ (\delta U\beta)^2],
  \end{equation}   
  where $\tilde \chi(\omega) = \int_0^{\infty} ds e^{i(\omega+i0 s)} \chi(s)$
  is the Fourier transform of linear response function
  \begin{equation}  
  \label{dd}
  \chi(s)= -i \frac{1}{Z}\TR \big( e^{-\beta H}  [\,e^{iHs}d e^{-iHs},d\,] \big)
  \end{equation}  
  of the final Hamiltonian $H(t>0)$. Equation (\ref{dqu}) holds for $\omega \neq 0$ 
  and $\tmax \to \infty$, while evaluation of $\tilde d(\omega)$ at $\tmax<0$ corresponds 
  to a broadening of $\chi(\omega)$ on the scale $1/\tmax$. The time-independent expansion 
  parameter $\beta \delta U$ in Eq.~(\ref{dqu}) is obtained by performing a perturbation 
  expansion of the initial state density matrix $\exp[-\beta H(t<0)t]$ instead of the 
  time-evolution operator $\exp[-iH(t>0)t]$. 
  
  In principle, $\chi(\omega)$ could be computed directly from equilibrium DMFT. However,
  the equilibrium imaginary time formalism requires an analytical continuation, while 
  the nonequilibrium calculation gives direct access to frequency or time-dependent 
  quantities. Furthermore, in experiment the values of $U$ would have to be changed 
  by at least a few per cent, such that it is unclear whether Eq.~(\ref{dqu}) is still 
  appropriate to describe the response. 
  
   \begin{figure}
  %\centerline{\includegraphics[width=1\columnwidth]{Figures/plots_neu/dquench_oca_b10.eps}}
  \centerline{\includegraphics[width=1\columnwidth]{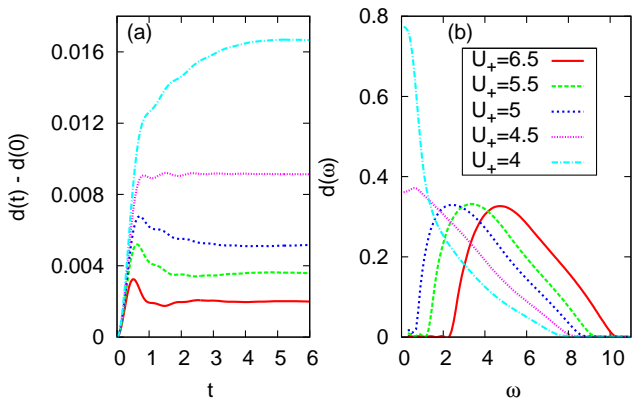}}
  \caption{
  (a) The double occupancy after quenches from $U=U_0$ to $U=U_0-\delta U$, with 
  $\delta U=0.5$, $\beta=10$, and $U_0=7,6,5.6,5,4.5$. OCA has been used as an impurity 
  solver.
  (b) Fourier transform (\ref{doccft}) of the date in (a).
  }
  \label{fig-dquench}
  \end{figure}

  After an interaction quench from the noninteracting ground state to the insulating phase
  in the Hubbard model, the double occupancy $d(t)$ relaxes through a series of oscillations 
  at approximate frequency $U$,\cite{Eckstein2009a} which correspond to the well-known collapse 
  and revival oscillations in the limit $U \gg V$.\cite{Greiner2002b} Similar oscillations become 
  visible after small quenches within the insulating phase ($U_+=6.5,5.5,5$ in Fig.~\ref{fig-dquench}); 
  their Fourier transform leads to a broad peak around $\omega\approx U$ (Fig.~\ref{fig-dquench}b). 
  The series of quenches displayed in Fig.~\ref{fig-dquench} corresponds to a scan through the 
  phase diagram at constant temperature $T=0.1$ above the critical temperature of the Mott 
  transition. A crossover between metallic and insulating behavior is thus expected between 
  $U\approx 4$ and $U \approx 5$. In agreement with this, the gap in $d(\omega)$ at $\omega=0$ 
  disappears between $U=5$ and $U=4.75$. Although the absolute changes of the double occupancy 
  are small, these results suggest that the double occupancy after an interaction quench may be 
  used as probe of the metal-to-insulator crossover in the Hubbard model.

  \section{Conclusion}
 
  In this paper we have presented a self-consistent diagrammatic strong-coupling expansion 
  on the Keldysh contour, which can be used to solve the Anderson impurity model in rather 
  general nonequilibrium situations. The main purpose of this study was the development 
  of an impurity solver for nonequilibrium DMFT which can cover the regime of large 
  interactions and relatively long times. By comparing results from our strong-coupling 
  expansion to numerically exact weak-coupling CTQMC for the Hubbard model, we found that 
  the strong-coupling expansion is a good candidate to fulfill these requirements: While 
  it fails in the metallic phase and at very low temperatures, and while the first order 
  or non-crossing approximation is correct only deep in the insulating phase, an important 
  correction arises already in the second order (OCA). In the insulating phase and in the 
  crossover regime the latter gives quite reliable results, which can be brought into 
  almost quantitative agreement with CTQMC by going to the third order of the expansion.
  
  Although the numerical effort for the evaluation of the diagrams rises considerably 
  with the expansion order, even calculations up to third order can be carried to 
  substantially longer times than CTQMC, because the latter suffers from an exponential 
  increase of the computational cost due to the dynamical 
  sign problem. We thus believe that the strong-coupling expansion will allow to extend 
  nonequilibrium DMFT investigations into the parameter regime of rather strong 
  interactions and not too low temperatures  which was so far not accessible 
  within weak-coupling CTQMC. It is precisely this parameter regime which is relevant for 
  many experiments with cold atomic gases\cite{Joerdens2008a,Strohmeier2010a} 
  and pump-probe spectroscopy. The broad range of possible applications of DMFT in this field 
  includes the excitation of the Mott insulating phase by a short laser pulse (similar 
  to the experiments in Refs.~\onlinecite{Perfetti2006a} and \onlinecite{Wall2009}), 
  the respose of the Mott insulator to very strong electrical fields that might lead 
  to a dielectric breakdown,\cite{Oka2003a} or an extended investigation of the 
  interaction quench in the Hubbard model.\cite{Eckstein2009a} 	

  In the last part of the paper we have used the self-consistent hybridization expansion 
  as an impurity solver within nonequilibrium DMFT in order to study the generation of 
  doubly occupied sites in a Mott insulator by a time-dependent variation of the interaction
  or hopping strength. In agreement with previous investigations,\cite{Kollath2006a,Huber2009a}
  the modulation spectrum was found to have a gap at $\omega=0$ and a pronounced maximum 
  at $\omega \approx U$. The maximum persists in the crossover regime, but its 
  location is then no longer proportional to $U$. Furthermore, we have studied the double 
  occupancy $d(t)$ as a function of time after small interaction quenches. In the crossover 
  regime, the behavior of $d(t)$ is drastically changing, which is most clearly evidenced 
  by the disappearence of the gap in the Fourier transform of $d(t)$. 
  Although the absolute changes of the double occupancy are small, its time-evolution 
  may thus yield a sensitive probe of the metal-to-insulator crossover in the Hubbard 
  model. 

  An interesting next step would now be to repeat similar calculations in 
  a more realistic setup, e.g., on a cubic lattice. A careful comparison of results 
  from the self-consistent hybridiation expansion up to second and third order, and 
  from CTQMC (for small times) will allow to make definite experimental predictions, 
  such as for the modulation spectrum in the insulating phase and the crossover 
  regime, and for the saturation behavior of the double occupancy. 
  
  \section*{Acknowledgements}
   
  This work was supported by the Swiss National Science Foundation  
  (Grant PP002-118866). CTQMC calculations were run on the Brutus  
  cluster at ETH Zurich, using the ALPS library.\cite{ALPS}

 \end{document}